\documentclass[aps,prc,twocolumn,nofootinbib]{revtex4-1}
\usepackage{amsmath,graphicx}

\begin{document}
\preprint{Saclay/T-10/...}

\title{Constraining the viscous freeze-out distribution function with data 
obtained at the BNL Relativistic Heavy Ion Collider (RHIC)}


\author{Matthew Luzum}
\author{Jean-Yves Ollitrault}
\affiliation{Institut de Physique Th\'eorique,\\
CEA, IPhT, F-91191 Gif-sur-Yvette, France\\
CNRS, URA 2306, F-91191 Gif-sur-Yvette, France} 
\date{\today}

\begin{abstract}
We investigate the form of the viscous correction to the equilibrium distribution function in the context of a Cooper-Frye freeze out prescription for viscous hydrodynamic simulations of relativistic heavy ion collisions.  The standard quadratic ansatz used by all groups for the case of shear viscosity is found to be disfavored by experimental data for $v_4/(v_2)^2$ at the Relativistic Heavy Ion Collider, and is unlikely to be correct for the hadron resonance gas present at freeze out.  Instead, data for $v_2(p_t)$ along with $v_4/(v_2)^2$ favor a momentum dependence between linear and quadratic.
\end{abstract}


\maketitle

\section{Introduction}
There has been much success in modeling the medium created in collisions at the Relativistic Heavy Ion Collider (RHIC) as a fluid.  First the equations of ideal hydrodynamics were used with surprising success, indicating that the collision fireball appeared to behave like a nearly perfect liquid \cite{Kolb:2001qz,Teaney:2000cw,Huovinen:2001cy,Hirano:2002ds,Kolb:2002ve}.  More recently, this ideal hydrodynamical description has been generalized to viscous hydrodynamics in an effort to quantify dissipative corrections to perfect fluid behavior\cite{Luzum:2008cw,Masui:2009pw,Drescher:2007cd,Heinz:2009cv,Bozek:2009dw}]  

The results of these simulations are, of course, sensitive to the initial conditions that are implemented for hydrodynamic evolution, which are a topic of much study.   However, the results can also be sensitive to the prescription used to describe the cessation of hydrodynamic behavior once the system has become dilute (``freeze out'').  This freeze out prescription must describe the spectrum of particles that comes from a particular piece of fluid when it decouples.  
In principle this distribution is obtained from a kinetic description of the system at freeze out, which for a viscous fluid is of the form $f = f_0 + \delta f$, where $f$ is the equilibrium distribution and $\delta f$ is a small correction.  

The exact form of $\delta f$, however, depends on the particle system in question.  As pointed out in Ref.~\cite{Dusling:2009df}, until recently (at least in the case of vanishing bulk viscosity), all groups have made the assumption that $\delta f \propto p^2 f_0$.  However, there is no particular reason to believe that this is the correct momentum dependence for the system of hadrons and resonances at freeze out in a relativistic heavy ion collision.  

In this article we use viscous hydrodynamic simulations to investigate the effect of using different forms for $\delta f$ on calculated observables in the absence of bulk viscosity and baryon chemical potential.  We find that experimental data, especially $v_4/(v_2)^2$, can be used to constrain $\delta f$, in principle giving information about the dynamics of the hadron resonance gas present at freeze out.   
%
%
%
%
\section{Viscous Hydrodynamics}
If a system is near thermal equilibrium, it is useful to write its energy-momentum tensor as a sum of a locally isotropic (ideal fluid) part $T_0^{\mu\nu}$, and a viscous correction $\Pi^{\mu\nu}$.  If, following Landau, we define the local rest frame as the zero momentum frame $T^{\mu0}=0$, we can write the result in terms of the four-velocity  $u^\mu$ of this local rest frame with respect to some fixed lab frame:
\begin{equation}
T^{\mu\nu} = T_0^{\mu\nu} + \Pi^{\mu\nu} =   \epsilon\ u^\mu u^\nu- p\ \Delta^{\mu \nu}  + \pi^{\mu\nu} + \Delta^{\mu\nu} \Pi\, .
\end{equation}
Here $\Delta^{\mu \nu} \equiv (g^{\mu \nu}-u^\mu u^\nu)$ is a projector onto the space orthogonal to $u^\mu$, while $\epsilon$ and $p$ are the energy density and pressure in the local rest frame.  
The viscous tensor $\Pi^{\mu\nu}$ is symmetric and transverse to the fluid velocity:   $u_\mu\Pi^{\mu\nu} = 0$.  
It is typically broken into a traceless shear tensor $\pi^{\mu\nu} \equiv (\Pi^{\mu\nu} - \frac 1 3 \Pi^{\alpha}_{\alpha} g^{\mu\nu})$ and a term proportional to the trace, the bulk pressure $\Pi \equiv \frac 1 3 \Pi^{\mu}_{\mu}$.  
To first order these terms are proportional to the shear viscosity $\eta$ and bulk viscosity $\zeta$, respectively:
%
\begin{align}
%
\pi^{\mu\nu} =&\  \eta\, \partial^{\langle\mu} u^{\nu\rangle} + O(\partial^2)\\
\Pi =&\ \zeta\, \partial_\alpha u^\alpha + O(\partial^2)\, . 
\label{pimunu}
%
%
\end{align}

The equations of viscous hydrodynamics then follow from the local conservation of energy and momentum
\begin{equation}
\partial_\mu T^{\mu\nu} = 0\, .
\end{equation}
Any other conserved quantity---such as baryon number---will have its own conservation equation and transport coefficients, but will be neglected here.
Along with the equation of state of the particular system in question $p = p(\epsilon)$ and the specified form of the viscous tensor (in principle including the temperature dependence of the transport coefficients), this uniquely specifies the evolution of a viscous fluid.
\section{Freeze out}
To model the breakdown of hydrodynamic behavior once a system has become cool and dilute, one must define a freeze out process.  This is done by taking a given fluid cell (once it meets some predefined criteria such as reaching a certain temperature $T_f$) and converting it into a distribution of particles $f$ given by kinetic theory.  In the standard Cooper-Frye prescription, these particles undergo no further interaction, and the finally observed particle spectrum is given by integrating this distribution over the freeze out surface $\Sigma^\mu$:
\begin{equation}
E\frac{d^3 N}{d^3 {\bf p}}\equiv 
\sum_i
\frac{d_i}{(2 \pi)^3} \int p_{\mu} d\Sigma^\mu f_i(p^\mu)\, ,
\end{equation}
where we have summed over the contribution from different particle species $i$ with degeneracy $d_i$.  Once thus obtained, it is often useful to express the angular dependence in terms of Fourier components, which for collisions of identical nuclei at midrapidity takes the form
\begin{equation}
E\frac{d^3 N}{d^3 {\bf p}}
= v_0 \left[1 + 2 v_2\, \cos(2\, \phi) +  2 v_4\, \cos(4\, \phi) + \ldots \right] .
\end{equation}

For a system close to equilibrium, the distribution function can be written as the equilibrium Bose/Fermi distribution $f_0$, plus a small viscous correction that vanishes in the ideal hydrodynamic limit
\begin{equation}
f(p^\mu) = f_0(p\cdot u) + \delta f(p^\mu)\, .
\end{equation}

The exact form of $\delta f$ depends on the dynamics of the particles in question \cite{Dusling:2009df}, so it will be useful to first derive the general properties that it must obey for any system.  For simplicity, consider a single species with unit degeneracy in the local rest frame of the fluid with $\delta f = \delta f({\bf p})$.

For consistency, the system of particles should generate the same energy-momentum tensor as the fluid just before freeze out:
\begin{equation}
T^{\mu\nu} = \int \frac {d^3p} {(2\pi)^3E} p^\mu p^\nu f({\bf p})\, ,
\end{equation}
which implies
\begin{equation}
\label{cont}
\Pi^{\mu\nu} = \int \frac {d^3p} {(2\pi)^3E} p^\mu p^\nu \delta f({\bf p})\, .
\end{equation}
The most general possible form, then, is:
\begin{align}
\delta f ({\bf p}) & =  \Pi_{ij} \left[ A(p) \left(\frac {1} {p^2} p^i p^j - \frac 1 3 \delta^{ij}\right) + \frac 1 3 B(p)  \delta^{ij} \right] \nonumber \\
& = A(p) \frac {1} {p^2} p^\mu p^\nu \pi_{\mu\nu} + B(p) \Pi\, .
\label{defAB}
\end{align}
Equation \eqref{cont} for the spatial components $\Pi^{ij}$ in the local rest frame gives a normalization condition for $A$ and $B$
\begin{align}
\label{norm}
\frac {2} {15} \int \frac {d^3p} {(2\pi)^3E} p^2 A(p) = 1&\\
\frac 1 3 \int \frac {d^3p} {(2\pi)^3E} p^2 B(p) = 1&
\end{align}
In the case of non-vanishing bulk pressure, Eq.~\eqref{cont} gives yet another constraint.  Recall that in the local rest frame we must also have $u_\mu \Pi^{\mu\nu} = \Pi^{0\mu} = 0$.  In particular, when we demand the Landau matching condition to be satisfied $u_\mu \Pi^{\mu\nu} u_\nu = \Pi^{00} = 0$, we find
\begin{equation}
\Pi \int \frac {d^3p} {(2\pi)^3}E\, B(p) = 0,
\end{equation}
which puts a nontrivial constraint on $B(p)$ when $\Pi \neq 0$.

Note that taking $B \propto p^2 f_0$ as in Ref.~\cite{Dusling:2007gi} fails to satisfy the Landau matching condition \cite{Monnai:2009ad}---in fact, this is the case for any positive definite function.  

Investigating $B(p)$ is left to future work.  In the following we will only concern ourselves with the case of vanishing bulk pressure, where $B(p)$ plays no role.

In contrast to $B(p)$, there is no constraint on the functional form
of $A(p)$, which is related to the more standard notation 
$\chi(p)$ by~\cite{Dusling:2009df} 
\begin{equation}
A(p)\equiv \frac 1 {2\eta} f_0(p)[1\pm f_0(p)]\chi(p).
\label{defchi}
\end{equation}
Until recently, all groups doing hydrodynamic simulations
have assumed $\chi \propto p^2$.  Although this is correct for the
most familiar solvable systems \cite{Dusling:2009df}, 
there is no particular reason to assume it to be correct at freeze out
for a heavy ion collision, and we will see that that it is in fact disfavored by data. 

%
%
%
\section{Microscopic perspective}
%
%
In principle $\delta f$ is given by a kinetic description of the particles comprising the system at freeze out---here a collection of hadrons and resonances not far below the deconfinement transition temperature.  In kinetic theory, the distribution function is determined by the Boltzmann equation
\begin{equation}
\label{Boltzmann}
(\partial_t + v_{\bf p} \cdot \partial_{\bf x})f({\bf p},{\bf x}) = -{\rm C}[f,{\bf p}]\ .\\
\end{equation}
In equilibrium the collision term ${\rm C}[f,{\bf p}] = 0$, and the distribution function is
\begin{equation}
f({\bf p},{\bf x}) = f_0(p,{\bf x}) = \frac {1} {e^{\,p_{\mu} u^{\mu}/T}  \mp 1}\, .
\end{equation}
By inserting $f = f_0 + \delta f$ into Eq.~\eqref{Boltzmann} and keeping only terms to first order in $\delta f$, one obtains the linearized Boltzmann equation.  For a vanishing bulk viscosity, in the local rest frame of the fluid, this is \cite{Dusling:2009df}:
\begin{equation}
\label{linboltz}
   \frac{p^{\mu}{p^\nu}}{E\, T} f_0(1 \pm f_0)\nabla_{\langle\mu} u_{\nu\rangle} 
 = -{\cal C}[\delta f,{\bf p}] \, ,
\end{equation}
where ${\cal C}[\delta f,{\bf p}]$ is the collision operator  expanded
to first order in $\delta f$.  
Using Eqs.~(\ref{pimunu}), (\ref{defAB}) and (\ref{defchi}), 
and neglecting quantum statistics, one obtains after some algebra:
\begin{align}
\label{linboltz2}
\frac{p}{T}&=\int \frac{d^3{\bf k}}{(2\pi)^3}v_{\rm rel}d\sigma_{{\bf pk}\to {\bf
    p'k'}}\\
\times
&\left(\chi(p)+P(\theta_{pk})\chi(k)-P(\theta_{pp'})\chi(p')-P(\theta_{pk'})\chi(k')\right),\nonumber
\end{align}
where $d\sigma_{{\bf pk}\to {\bf p'k'}}$ is the differential cross section, 
$P(\theta)=(3\cos^2\theta-1)/2$, 
and $v_{\rm rel}$ is the relative velocity:
\begin{equation}
v_{\rm rel}\equiv \frac{\sqrt{(p^\mu k_\mu)^2-(p^\mu p_\mu)(k^\nu k_\nu)}}{p^0 k^0}. 
\end{equation}
Eq.~(\ref{linboltz2}) is Eq.~(B5) of Ref.~\cite{Dusling:2009df}, with
different notation to emphasize that one may also think in terms of cross sections rather than energy loss rates. 
%

The relevant cross sections here are hadronic cross sections, since
freeze out occurs in the hadronic phase. Let us estimate the typical
energy of a hadronic collision at freeze out, which will indicate the relevant energy regime of these hadronic cross sections.
In the above equation, $p$ 
denotes the momentum of the hadron in the rest frame of the fluid. If
the hadron momentum is parallel to the fluid momentum, $p=u^0 p_t-u
m_t$, where $p_t$ and $m_t$ are the transverse momentum and transverse
mass in the lab frame, and $u^0=\sqrt{1+u^2}$. The fluid
velocity in realistic hydrodynamic simulations reaches values as large
as $u\sim 1$. An observed pion with $p_t$ in the range
1--4~GeV comes from a part of the freeze out surface that has a fluid velocity that is parallel to its momentum and is near this maximum $u$.  The momentum in the fluid frame $p$ is significantly
smaller---it is in the range 
0.4--1.7~GeV; a collision with a thermal pion at $T=0.14$~GeV
then occurs at $\sqrt{s}$ in the range 0.4--1~GeV. 
Hadronic cross sections have a nontrivial dependence on $\sqrt{s}$ in
this range. 
Ultimately, a consistent description of heavy-ion collisions using
viscous hydrodynamics should implement realistic hadronic
cross sections, but this is beyond the scope of the present paper;  the linearized Boltzmann equation is an integral equation that is nontrivial to solve even for relatively simple systems.

However, one can gain insight on how $\chi(p)$ depends on the cross section by
studying the limit where the momentum $p$ is much larger than 
the temperature $T$. This limit is relevant to the study of
anisotropic flow ($v_2$ and  $v_4$) at large transverse momentum $p_t$
(see Sec.~\ref{sec:results}). 
In this limit, 
outgoing momenta ${\bf p'}$ and ${\bf k'}$ are essentially
parallel to ${\bf p}$, so that $P(\theta_{pp'})\simeq
P(\theta_{pk'})\simeq 1$, and Eq.~(\ref{linboltz2}) simplifies to 
\begin{align}
\frac{p}{T}&=\int \frac{d^3{\bf k}}{(2\pi)^3}v_{\rm rel}d\sigma_{{\bf pk}\to {\bf
    p'k'}}\cr 
\times
&\left(\chi(p)+P(\theta_{pk})\chi(k)-\chi(p')-\chi(k')\right).
\label{linboltz3}
\end{align}
If $\chi(p)$ is a power law at large momentum $p$, $\chi(p)\propto
p^\alpha$ with $\alpha>0$, one can neglect $\chi(k)$ in the right-hand
side. The energy of the collision in the center of mass is of
order $\sqrt{s}\sim \sqrt{pT}$. 
By simple power counting, Eq.~(\ref{linboltz3}) then implies
that the cross section varies with $s$ like $s^{1-\alpha}$. 
The quadratic ansatz $\chi(p)\propto p^2$ solves the Boltzmann
equation at large momentum only if the cross section decreases like $1/s$. 
This is the case for, e.g,  $\phi^4$ theory~\cite{Dusling:2009df}, but not for
hadrons in the relevant range of $\sqrt{s}$, where hadronic cross
sections rather {\it increase\/} with $\sqrt{s}$. 
\section{Setup}
We would like to investigate the effect of using various forms for $\delta f$ at freeze out.  To do this, we implement a 2+1 dimensional conformal viscous hydrodynamic model with both Glauber- and CGC-type initial conditions, a constant value of $\eta/s$, a realistic equation of state, and a Cooper-Frye freeze out prescription including resonance feed down.  All parameters are taken from the best fit results from Ref.~\cite{Luzum:2008cw}, to which we refer the reader for all relevant details.  The only change here is to replace the quadratic ansatz for $\delta f$ ($\chi = c\, p^2$) with two other simple choices:  a linear ansatz ($\chi = c\, p$), and a power law in between linear and quadratic ($\chi = c\, p^{1.5}$).  It has been argued \cite{Dusling:2009df} that the correct momentum dependence for $\delta f$ is expected to lie somewhere within this range.    For each case, the same form of $\delta f$ is used for all particles at freeze out (using a Boltzmann equilibrium distribution, ignoring quantum statistics), with the overall temperature dependent normalization $c$ then fixed by the generalization of Eq.~\eqref{norm}:
\begin{equation}
\label{norm2}
\frac {2} {15} \sum_i d_i \int \frac {d^3p} {(2\pi)^3E_i} p^2 f_0(p)\frac {\chi(p)}{2\eta} = 1.
\end{equation}
Note that, in principle, even with the same momentum dependence, different particle species in a particular system could have different size contributions to the shear tensor (see, e.g., Sec. IV of Ref.~\cite{Dusling:2009df}), but this possibility is not investigated here.

All hadrons and resonances up to $\sim$2 GeV are calculated, with unstable resonances then allowed to decay after freeze out.
%
%
%
%
%
%
\section{Results}
\label{sec:results}
\begin{figure*}
\includegraphics[width=.497\linewidth]{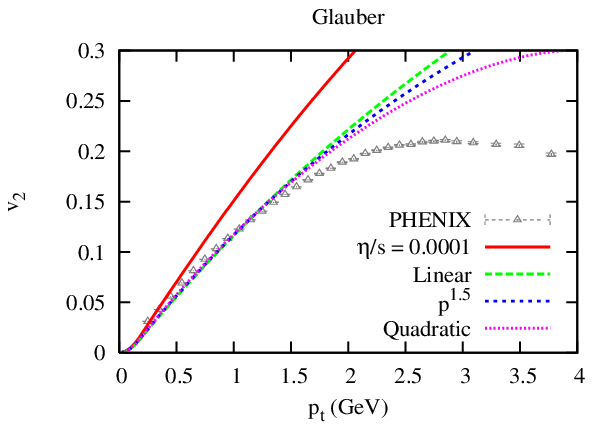}
\includegraphics[width=.497\linewidth]{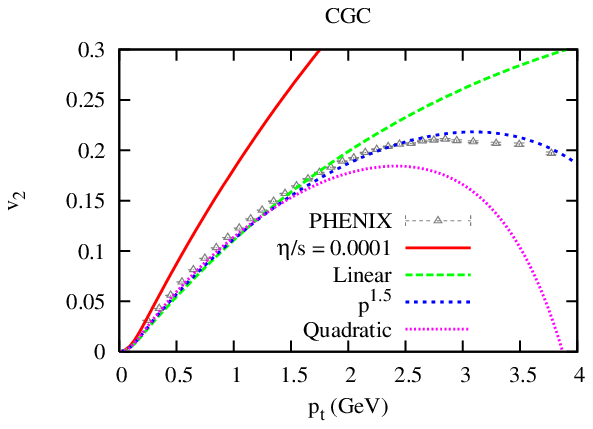}\\
\includegraphics[width=.497\linewidth]{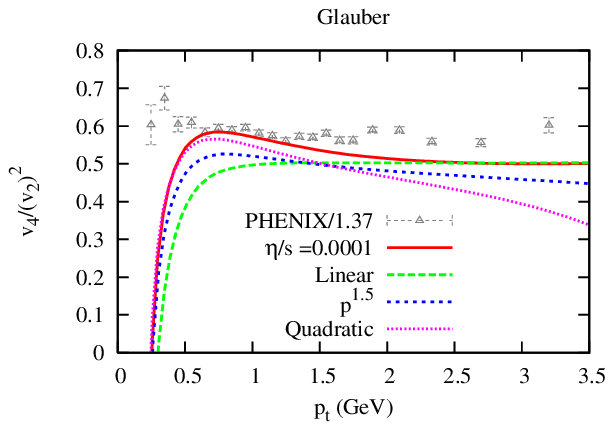}
\includegraphics[width=.497\linewidth]{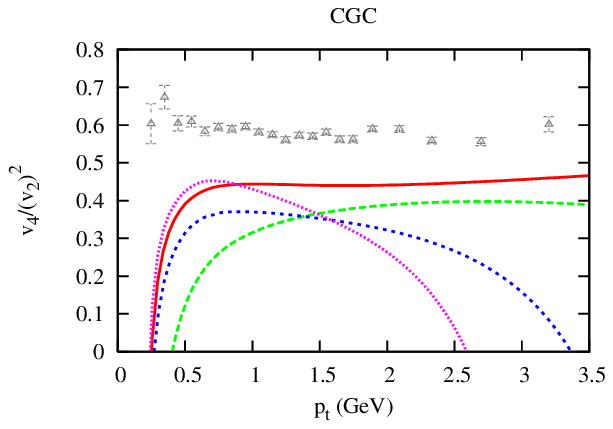}\\
\caption{(Color online) Charged hadron $v_2$ and $v_4/(v_2)^2$ at impact parameter $b$ = 8 fm for three choices of $\delta f$: Linear ($\chi \propto p$), Quadratic ($\chi \propto p^2$), and in between ($\chi\propto p^{1.5}$), with ideal hydrodynamics results included for comparison.  The viscous Glauber calculation is for $\eta/s = 0.08$ and $T_i = 0.333$ while the CGC calculation used $\eta/s = 0.16$ and $T_i = 0.299$ \cite{Luzum:2008cw}.  Experimental data for $v_4/(v_2)^2$ scaled down by a factor of 1.37 to account for fluctuations \cite{Gombeaud:2009ye, Luzum:2010ae}.  Data at 30-35\% centrality from PHENIX with statistical error bars only \cite{ :2010ux}. }
\label{v4}
\end{figure*}
Given the restriction of Eq.~\eqref{norm}, any observable that is integrated over transverse momentum will be largely insensitive to the choice for $\delta f$.  We found this to be true in our calculations for, e.g., $\langle p_t \rangle$ and multiplicity.  However, the momentum dependence of differential quantities such as $v_2(p_t)$ and $v_4(p_t)$ can strongly depend on this choice.  Figure \ref{v4} shows differential $v_2$ and $v_4/(v_2)^2$ for charged hadrons from our viscous hydrodynamic calculations as compared to experimental data from PHENIX at RHIC.  As was found previously~\cite{Dusling:2009df}, $v_2$ below $p_t\sim1.5$ GeV is largely unchanged for a range of reasonable choices for $\delta f$.  At larger $p_t$, however, it can change the shape of the curve significantly.  A stronger momentum dependence results in a larger viscous correction---and thus a larger deviation from ideal hydrodynamic behavior---at larger momentum.  In contrast, $v_4/(v_2)^2$ depends strongly on the choice for $\delta f$ even at smaller momentum.

As can be seen in the plots for $v_2$ in Fig.~\ref{v4},  the shape of the $v_2$ curve depends on a combination of viscosity and the choice for $\delta f$, while the normalization depends mostly on initial eccentricity and viscosity.  Experimental data appear to favor a comparatively large initial eccentricity, such as can accommodate a viscosity of order $\eta/s\sim0.16$, while the favored $\delta f$ depends on this viscosity value.  A linear ansatz seems unlikely, while a stronger momentum dependence works well.

The curve for $v_4/(v_2)^2$, on the other hand, favors a weak momentum dependence.  The quadratic ansatz in particular is quite difficult to reconcile with the very flat experimental data---at least for $\eta/s > 0.08$ and/or for more peripheral collisions.  A $\delta f$ with a stronger momentum dependence also results in a strong impact parameter dependence, which is not seen in data (see Ref.~\cite{Luzum:2010ae}), while a linear ansatz results in a dependence on impact parameter that is even smaller than is seen from ideal hydrodynamics.  

%
Note that for any choice of $\delta f$, the magnitude of 
$v_4/(v_2)^2$ is somewhat smaller than the experimental data, even after 
correcting for the estimated effect of eccentricity fluctuations and 
especially for CGC-type initial conditions.  This could be due to the 
presence of additional fluctuations or differences in the initial 
energy density profile, and this is not yet fully understood.  The 
shape of the curve as a function of $p_t$, however, is not 
sensitive to these aspects of the calculation and is largely 
determined by the viscous correction (in conjunction with the freeze 
out temperature)~\cite{Luzum:2010ae}.  In addition, both STAR and PHENIX find a flat 
$p_t$ dependence, though at different values.  It therefore appears that 
systematic errors in the experimental data may affect the magnitude of 
$v_4/(v_2)^2$, but are unlikely to strongly affect the shape. Thus, it 
is mostly the shape of the curve that matters here, and this is what 
provides the constraint on the viscous correction, clearly showing a 
preference for a momentum dependence that is weaker than quadratic.

Thus, data as well as theoretical considerations point to something between the linear and quadratic ansatz.  

One should refrain from trying to extract a precise value from the finer details of Fig.~\ref{v4}, however, since certain details could change with an improved treatment.  For example, our model uses a simple Cooper-Frye freeze out prescription with simultaneous chemical and kinetic freeze out, and the initial conditions are still quite uncertain (hence the use of two different types of initial conditions), and so not all the physics is perfectly captured.  Even within our model, an interplay with freeze out temperature could change which exact $\delta f$ is most favored by $v_4/(v_2)^2$ data \cite{Luzum:2010ae}.   In addition, hydrodynamic results in general have uncertain applicability at large $p_t$.  As $p_t$ increases, there is an increased probability that other physics is coming into play and one should have less confidence in hydrodynamic calculations.

Nevertheless, it is clear that  the uncertainty involved in the correct form for $\delta f$ has potentially measurable effects.  Further improvements in hydrodynamic calculations could therefore potentially give valuable information about the microscopic dynamics at freeze out by comparing to data.  Conversely, theoretical progress from the microscopic side could give information about where the validity of the hydrodynamic description starts to break down.
\section{Conclusion}
It has been thus far unclear what is the correct form for the viscous correction to the equilibrium distribution function used at freeze out in viscous hydrodynamic calculations of heavy ion collisions.  We investigated the effect of various reasonable choices for $\delta f$ in addition to the standard quadratic ansatz on the resulting differential elliptic flow $v_2 (p_t)$ as well as the higher-order harmonic $v_4 (p_t)$ and compared to experimental data.  The quadratic ansatz is strongly disfavored by RHIC data, with the most likely form lying somewhere between linear and quadratic.  Specifically, data for $v_4/(v_2)^2$ favor a weak momentum dependence in $\delta f$ both from the shape of the curve as a function of $p_t$, as well as the centrality dependence.  Data for $v_2(p_t)$ favor a stronger-than-linear momentum dependence in $\delta f$, but the precise best-fit choice is viscosity dependent.
\begin{acknowledgments}
We would like to thank Paul Romatschke for the hydrodynamics code.  This work was funded by ``Agence Nationale de la Recherche'' under grant
ANR-08-BLAN-0093-01.
\end{acknowledgments}


\end{document}